# Deformation twins as a probe for tribologically induced stress states

Antje Dollmann[1,2], Christian Kübel[3,4,5], Vahid Tavakolli[3,4,5], Stefan J. Eder[6,7], Michael Feuerbacher[8], Tim Liening[8], Alexander Kauffmann[1], Julia Rau[1,2,9], Christian Greiner[1,2]

[1] Institute for Applied Materials (IAM), Karlsruhe Institute of Technology (KIT), Kaiserstraße 12, 76131 Karlsruhe, Germany

[2] IAM-ZM MicroTribology Center (µTC), Straße am Forum 5, 76131 Karlsruhe, Germany

[3] Institute of Nanotechnology (INT), Karlsruhe Institute of Technology (KIT), 76344 Eggenstein-Leopoldshafen, Germany

[4] Institute of Materials Science, Technical University Darmstadt (TUD), 64287 Darmstadt, Germany

[5] Karlsruhe Nano Micro Facility (KNMFi), Karlsruhe Institute of Technology (KIT), 76344 Eggenstein-Leopoldshafen, Germany

[6] AC2T research GmbH, Viktor-Kaplan-Straße 2/C, 2700, Wiener Neustadt, Austria

[7] TU Wien, Institute of Engineering Design and Product Development, Lehárgasse 6, Objekt 7, 1060, Vienna, Austria

[8] Peter Grünberg Institut PGI-5 and ER-C, FZ Jülich GmbH, D-52425, Jülich, Germany

[9] Department of Physics, Chalmers University of Technology, 412 96 Göteborg, Sweden

## Abstract

Friction and wear of metals are critically influenced by the microstructures of the bodies constituting the tribological contact. Understanding the microstructural evolution taking place over the lifetime of a tribological system therefore is crucial for strategically designing tribological systems with tailored friction and wear properties. Here, we focus on single-crystalline CoCrFeMnNi that is prone to form twins at room temperature. Deformation twins feature a pronounced orientation dependence with a tension-compression anisotropy, a distinct strain release in an extended volume and robust onset stresses. This makes deformation twinning an ideal probe to experimentally investigate the complex stress fields occurring in a tribological contact. Our results clearly show a grain orientation dependence of twinning under tribological load. Unexpectedly, neither the crystal direction parallel to the sliding nor the normal direction are solely decisive for twinning. This experimental approach is ideal to experimentally validate tribological stress field models, as is demonstrates here.

# Introduction

Reductions in friction and wear have more than significant potential for the decrease of global $CO_2$ emissions[1]. This is highlighted by the fact that for example in the transportation sector 26 % of the invested primary energy is wasted to overcome friction forces and to generate unwanted wear[1]. Realizing this $CO_2$ reduction potential hidden in tribological contacts, requires fundamental knowledge about the origin of high or low friction and wear. In this manuscript, the focus lies on the microstructural evolution in metals and alloys, as the microstructure, e.g. grain size and orientation, has been known for decades to change under tribological loading[2,3]. It has been frequently observed that single-crystalline or coarse-grained metals exhibit grain-refinement under sliding[4–6], which influences the tribological properties[7]. The number of parameters influencing the microstructural evolution under tribological loading is large. Among the parameters are the crystallographic orientation[8–10], the material's stacking fault energy[11–13], temperature[14–16], sliding velocity[17–19] and the material pairing[20]. Regardless of all these parameters, the elementary processes causing the microstructure evolution are not easily predictable.

Various attempts have been made to understand the microstructural evolution in metals and alloys under tribological loading. The first observation concerning the microstructural evolution were the formation of subgrains and crystal rotation[5,21]. The crystal was found to preferably rotate around the axes perpendicular to the sliding and normal direction[5,22], but also other rotation axes have been observed[22–24]. For subgrain formation, first a grain boundary parallel to the surface (often called dislocation trace line (DTL) in literature) forms[25,26]. A model for its origin describes the glide of dislocations first into the bulk and afterwards towards the surface until the dislocations get stuck in the material as the critical resolved shear stress is no longer reached[27], while a more recent study claims a dislocation interaction[28]. Dislocation motion not only results in the formation of a DTL, but also in simple shear and crystal rotation as shown in a study in the vicinity of an annealing twin boundary[29]. The DTL formation was observed under various loading states and in different materials at the onset of the microstructural evolution[11,26,30]. As sliding continues, short grain boundary segments perpendicular to the surface form, which act as bridges between the elongated grain boundaries parallel to the surface and confine the subgrains[31]. With increasing cycle number, the grain size decreases further[25]. Calculations of the grain sizes have been conducted using a crystal plasticity approach[26] and via molecular dynamics (MD) simulations[13]. The challenge with dislocation-based plasticity is to analyse the active slip systems. Nevertheless, several groups have shown different procedures, which rely on slip trace analyses, calculations, or simulations. Slip traces are easily accessible by top-down imaging of the wear track[8,32]. The next level of complexity includes the analysis of slip traces from experiments on microwalls, which is experimentally much more challenging. There, the focus is on the slip systems underneath the indenter[28]. Note that for all slip trace analyses, only the slip plane and not the Burgers vector can be analysed. A combination of experiments and calculations is given by Cai et al.[23]. Based on the measured crystal orientation and the assumption of a simple shear stress field, Schmid factors on the slip systems were calculated. Discrete dislocation dynamics simulation revealed which dislocations can follow an indenter[33]. The most recent attempt was undertaken by a crystal plasticity finite element method, where Schmid factors and the accumulated shear strain on the slip systems[34,35] as well as crystal rotation[36] were calculated.

Despite all these impressive approaches, the prediction of microstructural evolution still remains a challenge. This is to a good part due to the fact that there exists only very limited knowledge about the stress field acting during the sliding contact. The most widely used stress field is that derived by Hamilton and based on Hertzian contact theory[37]. Understanding the stress is however an absolute necessity, if one wants to interpret the microstructural features as they present themselves in the post-mortem analyses highlighted above. As a model-based approach to the stress field is very challenging, experimental evaluations are basically the only avenue left. In case such an experimental approach were successful, this would have the added benefit that these experimental data could then be used to validate current and future modelling results. This directly leads to the question how to experimentally probe the stress field in a tribological contact. One would need a feature that is clearly recognizable in a micrograph and about which unambiguous statements can be made. Materials Science luckily knows such a feature, which – under the right conditions – can be active: Deformation twins. Deformation twins are unique as they appear on the planes with the highest resolved shear stress[38] and exhibit robust onset stresses. We therefore make use of this fact to gain insights into the tribological stress field. Deformation twins beneficially differentiate from dislocation motion: (1) twins are 3D extended defects with distinct strain release, which simplifies their detection, so they can be used as probes; (2) twin formation is dependent on the crystallographic orientation and exhibits a pronounced tension-compression anisotropy. To further reduce the complexity of the analysis, single crystal experiments are employed. Via a combination of crystal directions in sliding and normal direction, conclusions can be drawn on the decisive stress components for twin activation. Until now, deformation twinning has garnered only little interest in the tribological community[9,39,40]. Based on our previous results for polycrystalline samples[20,41], we formulate the following hypothesis: For a low friction coefficient, the normal direction is decisive for twinning, while for a high friction coefficient, the sliding direction is decisive[20]. We here experimentally put this hypothesis to the text and additionally employ MD simulations and calculations of resolved shear stresses based on the Hamilton stress field.

## Results & Discussion

All tribological experiments were conducted on equiatomic CoCrFeMnNi single crystals. This material was chosen as it is known to twin at room temperature under tribological loading[20]. It has a stacking fault energy of (30 ± 5) mJ/m$^2$ [42]. The experiments were conducted with a normal load of 2 N acting on a spherical counter body made out of SiC (hightech ceram, Dr. Steinmann + Partner GmbH, Germany) with a diameter of 10 mm without lubrication and in air (50 % RH). This material pairing is known to result in a friction coefficient of around 0.2[20], which is low for a dry contact. We here measured the same friction coefficient and for different crystal orientations (Figure S1). A single-trace protocol was chosen to facilitate assigning a microstructural feature to a stress component of the tribological stress field. The sliding distance was 6 mm. For the given approach, the crystallographic directions parallel to the sliding direction (SD) and the normal direction (ND) are crucial parameters. As mentioned above, twinning is dependent on the crystal direction parallel to the loading axis and the sign of the applied uniaxial stress[43]. Crystal directions within the [001]-[012]-[113] sector of an inverse pole figure of the loading axis are known for twin formation in compression, while crystal directions parallel to the loading axis within the region [012]-[011]-[111]-[113] tend to exhibit deformation twinning in tension[43]. Here, crystal directions that exhibit deformation twinning under these simplified conditions of uniaxial loading in compression or tension were combined systematically with each other. This resulted in the following couples for the normal and sliding directions of the crystals in question: (1) ND $[00\bar{1}]$ SD $[100]$, (2) ND $[00\bar{1}]$ SD $[\bar{1}10]$, (3) ND $[0\bar{1}\bar{1}]$ SD $[100]$, and (5) ND $[0\bar{1}\bar{1}]$ SD $[0\bar{1}1]$. Additionally, one experiment on a single crystal with (5) ND $[0\bar{1}\bar{1}]$ was conducted in Shockley partial direction (SD $[\bar{2}\bar{1}1]$).

For all five wear tracks, STEM foils were cut in the centre of the wear track parallel to the SD. The corresponding STEM images are given in Figure *1*a-e. In all images, horizontal lines are visible. They occur at different depths. Transmission Kikuchi diffraction (TKD) measurements (Figure S2) identify these lines as either low angle (< 15°), high angle grain boundaries (> 15°) or twin boundaries. Horizontal grain boundaries, called dislocation trace lines (DTLs), have frequently been observed in literature under tribological loading in the early stages of microstructural evolution for varying materials and loading conditions[25,26,28]. The DTL is not further discussed, as it is not the focus of this manuscript. The STEM images in Figure *1*b+c with ND $[00\bar{1}]$ SD $[\bar{1}10]$ and ND $[0\bar{1}\bar{1}]$ SD $[100]$, respectively, show further lines tilted in SD underneath the DTL. As the resolution of the TKD measurements is not sufficient for their characterisation, high resolution transmission electron microscopy images were taken, and fast Fourier transforms (FFT) were computed. These images are presented in Figure *1*f+g. Twins result in additional diffraction spots in the FFT[44] besides the matrix spots. These spots were detected in the FFT for the STEM foil ND $[0\bar{1}\bar{1}]$ SD $[100]$ in Figure *1*f. Therefore, the line-type feature is unambiguously identified as a deformation twin. In the FFT in ND $[00\bar{1}]$ SD $[\bar{1}10]$ in Figure *1*g, only the matrix spots and streaking are observed. Streaking is a clear indicator for planar defects such as stacking faults[44]. Stacking faults form at low strains in twinning induced plasticity (TWIP) materials such as CoCrFeMnNi[45,46]. Most probably, twins in TWIP materials do not form via a pole mechanism[47] but via continuous stacking-fault mediated processes, resulting in imperfect twins[46,48,49]. The line-type features in Figure *1*g are, thus, stacking fault aggregates and therefore imperfect twins or precursors of twins. For both defects – stacking faults and twins, $\frac{1}{6}\langle 11\bar{2}\rangle$ Shockley partial dislocations glide on $\{111\}$ planes, meaning that the same slip systems are probed. To simplify the following discussion, both line-type features will be referred to as twins.

These results clearly show that also under tribological loading, the activation of twinning is dependent on the crystal orientation, which is in good agreement with another study[9]. There, the crystal orientation dependence of twinning under sliding conditions was investigated by ramping the normal load. The following crystal orientations in austenitic stainless steel were employed: ND $[010]$ SD $[101]$, ND $[010]$ SD $[001]$, and ND $[111]$ SD $[\bar{1}10]$. Only the last crystal orientation did not exhibit deformation twinning. These result of observed twins in ND $[010]$ SD $[001]$ is in contrast to our work, where no twins were observed in the crystal orientation of ND $[00\bar{1}]$ SD $[100]$.

This might be rationalized by the different experimental set-ups. Results from literature with only spherical indentation and without sliding will be considered for a first interpretation: In this case, twins were observed in grains with ND $[001]$, and no twins were observed with ND $[011]$[50,51]. This is consistent with the result that twins form with a loading axis parallel to $[001]$ under uniaxial compression and do not form with a loading axis of $[011]$[43,52]. Reiterating the above stated hypothesis derived from polycrystal experiments, the normal stress parallel to ND ($\sigma_{\mathrm{ND}}$) is decisive for twinning at the friction coefficient measured here. This means that indentation and tribological experiments result in the same onset of twinning. This is true for the study of Patil et al.[9], but not for ours, as in ND $[00\bar{1}]$ SD $[100]$ no twins were observed. The increasing normal load in the study of Patil et al. is more similar to indentation than sliding with a constant normal load. Therefore, it is not surprising that our results differ. The occurrence of twinning during tribological loading with constant normal load in Figure *1* has no preference concerning ND or SD. Hence, a conclusive criterion for twinning under tribological loading seems to remain open. This will be discussed in the following using different avenues.

First, the twin systems will be geometrically investigated. Both activated twin systems are tilted in SD. All microstructures exhibiting twinning under tribological load have at least one twin system tilted in the same way in literature[9,20,53]. This leads to the conclusion that $\{111\}$ planes tilted around TD in SD are favoured for deformation twinning. The slip pattern on tribologically loaded microwalls support this finding by having mainly slip traces that are also tilted around TD in SD[28]. The trace of the twin in the STEM images and the knowledge about SD and ND allow the determination of the twin planes. The knowledge of a twin plane limits the number of possible shear directions to three. Accessing the active shear direction is experimentally challenging. However, to assume that the shear direction featuring the smallest angle with the SD-ND plane is activated seems very reasonable, as this one can most easily be activated by the shear component of the stress field. Such an analysis results in a $(\bar{1}11)[\bar{1}1\bar{2}]$ twin system for ND $[00\bar{1}]$ SD $[\bar{1}10]$ and a $(\bar{1}\bar{1}\bar{1})[2\bar{1}\bar{1}]$ twin system for ND $[0\bar{1}\bar{1}]$ SD $[100]$. It is noted that the distinct indexing $(hkl)[uvw]$ (twin plane and shear direction) with the twin plane normal $[hkl]$ (due to a cubic crystal system) is relevant to define the direction of strain release by the twin. Both aforementioned systems are also marked in Figure *1*b+c. All four vectors – twin plane normal and Shockley partial dislocations – lie within the SD-ND plane. The other crystal orientations, in which no twinning was observed, do not contain twin systems with a possible twin plane normal and shear direction within the SD-ND plane. This seems to be a geometrical requirement for twinning under tribological load at constant normal load. Both crystal orientations that exhibit twinning under tribological loading have in common that their transverse direction (TD) is parallel to $\langle 110 \rangle$. All crystals having TD parallel to $\langle 110 \rangle$ have a $\{111\}$-plane perpendicular to the ND-SD plane and a shear direction for twinning within this plane. Therefore, we conclude that crystal orientations with $\langle 110 \rangle$ parallel to TD form twins.

A closer look at the marked twin systems in Figure *1*b+c reveals differences. In both cases, the shear direction (Burgers vector of the Shockley partial dislocation) is oriented towards the surface, but the twin plane normal oriented in opposite directions. In contrast to full dislocations, a reversal of the shear direction is not possible for twinning, because this causes an AA (as opposed to ABČBA) stacking sequence of the close-packed planes[38]. The variance in the twin systems gives a first hint regarding the differences in twin formation.

Two further methods were applied to understand the fundamentals for twin activation under tribological load. The first method are MD simulations, which are chosen for reflecting plasticity on an atomic level and for considering the rearrangement of atoms most localized in the subsurface region. The second method relies on the calculation of the macroscopic resolved shear stresses on the twin systems. For this, the linear elastic but inhomogeneous stress field approximation according to Hamilton[37] is used, which combines the Hertzian pressure with a shear stress proportional to the experimentally determined friction coefficient. Both methods will be discussed with regard to the experimentally chosen crystal orientations.

A sliding process of a spherical indenter on a single crystal with ND $[00\bar{1}]$ SD $[\bar{1}10]$ was performed using MD simulations, with the results being analysed with OVITO[54]. Two consecutive timesteps of the resulting microstructures are shown in Figure *2*a+b. The green dots are atoms with face centred cubic coordination (ABCABC stacking sequence of closed-packed planes) and the red ones are atoms with hexagonally closed-packed (hcp) coordination (ABABAB stacking sequence). Double layers with hcp atoms are intrinsic stacking faults (ABˇAB stacking sequence), and single layers are twin boundaries (ABČBA stacking sequence). Within the microstructure, obviously more than one twin system is activated. This is in part due to the manageable system sizes in MD simulations, which are usually limited to some tens of millions of atoms, while sliding velocities are typically in the range of single digit m/s and above. It is difficult to make any judgement about the size effect in our study, but it is well known from literature that the probability for twin activation increases with increasing strain rate[52]. Furthermore, this MD simulation does not take any surface roughness, chemical and/or adhesive interaction between the two bodies, or any explicit atmosphere into account. Nevertheless, these MD simulations are highly valuable as they can display the time-resolved atomic displacements of the experimentally observed twin in the insets of Figure *2*a+b and thereby allow us to understand the observed twin system.

The experimentally observed twin is underneath the indenter towards the leading edge (in SD). In case one aims at understanding the normal stress responsible for the activation of a deformation twin, one needs to know the atomic displacements of the twin itself. To our knowledge, only MD simulations offer the possibility to obtain this kind of information. As shown in Figure *2*a, one atom on each side of the twin was marked, and their distances parallel to SD and ND were calculated from the atomic positions in the MD simulation. In Figure *2*b, the atoms with the same IDs were marked after twin lath thickening (perpendicular to the twin plane) to accommodate strain, and their distances determined. Comparing the distances demonstrates that the distance decreases in ND but increases in SD. A twin schematic with the same atomic displacements is given in Figure *2*c. The twin consequently forms under compression in ND. The atomic displacement and the position of the twin – underneath the indenter – correspond to a compressive $\sigma_{\text{ND}}$ being responsible for twin activation with ND $[00\bar{1}]$ SD $[\bar{1}10]$. As discussed before, this is also in accordance with the results of spherical indentation tests[50,51] on austenitic steel on a single crystal with ND $[00\bar{1}]$.

The second method is a resolved shear stress (RSS) analysis on the slip and twin systems using the Hamilton stress field. It has to be mentioned that beyond the onset of plasticity a linear elastic stress field model overestimates the stresses, in some cases significantly. This overestimation can be easily understood by imagining Hooke's law and a real stress strain curve. In the plastic regime after the yield point, the extrapolation of Hooke's law always yields higher stresses for a specific total or equivalent strain than the stress-strain curve featuring plastic flow. Having said this, smaller effective depths of stress and strain field are to be anticipated under elastic-plastic than purely elastic conditions as they are assumed in Hamilton's approximation. A further limitation of the Hamilton stress field is its static nature. As mentioned above, the microstructural deformation behaviour may change with the strain rate. Other parameters that are not covered are the real contact area and the interaction between the bodies in contact. For the RSS analyses, the twin systems have to be defined first. Based on the atomic arrangement, only one shear direction is possible with a fixed twin plane normal to form a deformation twin[55]. All twin systems are defined that a positive RSS on the system activates deformation twinning. This analysis results in RSS values dependent on the lateral position and depth underneath the spherical indenter. Out of these stress distribution maps, the maximum RSS value is extracted. This was performed for all twelve twin systems and all possible SDs were captured around ND up to 360°. The corresponding rotation angle is designated $\varphi$. The maximum RSS values of the twelve twin systems as a function of $\varphi$ are presented in Figure *3*a. $\varphi = 0°$ describes the SD $[100]$ and $\varphi = 225°$ SD $[\bar{1}10]$, both for ND $[00\bar{1}]$. Apart from deformation twinning, dislocation activity needs to be assessed: (1) in order to form twins, certain dislocation activity is needed irrespective of the distinct twinning mechanism and (2) strain hardening might be needed to achieve sufficiently high internal stresses to reach the critical stress for deformation twinning. Thus, the RSS for all slip systems was calculated in the same way as for the twin systems. Note that there is no tension-compression anisotropy for slip and the absolute magnitude of the RSS is considered. In Figure *3*a, only the enveloping curve with the maximum RSS on the twelve slip systems is presented for better clarity of the diagram. In Figure S3a+c, all curves of the maximum RSS on the slip and twin systems are presented. At $\varphi = 225°$ (SD $[\bar{1}10]$), there is one of four global maxima in RSS of 226 MPa on the twin system $(\bar{1}11)[\bar{1}1\bar{2}]$. This is the same twin system as the geometrically analysed one for ND $[00\bar{1}]$ SD $[\bar{1}10]$, clearly indicating very good agreement between experiment and calculation. The critical RSS for twinning in single-crystalline CoCrFeMnNi has previously been determined to be in the range from 110 to 378 MPa[56–60]. The calculated 226 MPa thus lie within the expected range. For SD $[\bar{1}10]$, dislocation-based plasticity is additionally limited because of the minimum in the RSS on the slip systems and strain accommodation by twinning is reasonable.

For $\varphi$ = 0° (SD $[100]$), no twins were observed experimentally. Here, a minimum is observed in the RSS on the twin systems that lies marginally below the maximum at SD $[\bar{1}10]$. Considering the above-mentioned limitations of this calculation, it stands to reason that such a small difference of only 4

MPa might be sufficient for not obtaining twin formation. At the same time, the RSS for slip is considerably higher close to the maximum and thus strain accommodation by slip might reduce the geometrical necessity for twinning.

The MD simulations had strongly indicated that twins form mainly by a compressive normal stress parallel to ND in SD [$\bar{1}10$]. Therefore, the RSS on the twin systems were calculated based on a single stress component of the stress tensor mimicking this situation. Representative RSS distribution maps for $\sigma_{\text{ND}}$, the normal stress parallel to SD ($\sigma_{\text{SD}}$) and the shear stress in SD-ND ($\tau_{\text{SD-ND}}$) are presented in Figure *3*b+c for two exemplary twin systems. The first one is the twin system ($\bar{1}11$)[$\bar{1}1\bar{2}$] which was experimentally verified. The second system ($\bar{1}1\bar{1}$)[$\bar{1}12$] was chosen since the twin plane normal and the shear direction also lie within the SD-ND plane, but the tilt around TD is in contrast to the activated system against SD. All shear stress distribution maps of the entire stress field as well as the relevant stress components of the four systems with the highest resolved shear stresses (Figure *3*a) can be found in Figure S4. In Figure *3*b+c, the RSS based on $\sigma_{\text{ND}}$ is clearly the highest. The stress distribution maps for both twin systems are identical. The stress distribution map based on $\sigma_{\text{SD}}$ is also identical for both twin systems. This is not surprising, as both twin systems have the same inclination angle with respect to the described normal stresses. A difference can be observed in the stress distribution map for $\tau_{\text{SD-ND}}$. Here, the colours are inverted between the two investigated systems. This clearly shows that for such twin systems, $\tau_{\text{SD-ND}}$ is the decisive criterion determining which twin system is activated. Summing up the analyses for the single crystal with ND [$00\bar{1}$], the RSS calculations with the Hamilton stress field is in good agreement with the experimentally observed microstructure. Concerning twinning, both methods – the MD simulation and the RSS analysis – strongly indicate that the normal stress parallel to ND dominates the RSS on the twin systems. The other stress components are responsible for selecting the activated twin system. The twins form underneath the spherical indenter at the leading edge.

Similar analyses were conducted for the single crystal with ND [$0\bar{1}\bar{1}$]. Snapshots of the MD simulations with SD [$100$] are presented in Figure *4*a+b. The simulation results in more active twin systems than the experiment, as observed before. For these data, the experimentally observed twin will be further investigated in the MD simulation results, using the same methods as for ND [$00\bar{1}$] SD [$\bar{1}10$]. The first finding is the location of twin formation relative to the spherical indenter, which is significantly in front of the indenter. Next, we aim at identifying a possible normal stress leading to twinning. Distances in SD and ND between two atoms on the left and right side of a twin were determined in Figure *4*a+b.. These distances increase in vertical direction and decrease in horizontal direction. As visualized by the twin schematic in Figure *4*c, a twin with the given displacements can be formed by a compressive $\sigma_{\text{SD}}$. This constitutes another difference to the twin in ND [$00\bar{1}$] SD [$\bar{1}10$], where $\sigma_{\text{ND}}$ was decisive. This analysis showcases the strength and value of MD simulations. Although the microstructures found in the experiments are not perfectly replicated , it can very successfully be used to understand the basic mechanisms. The arrangement of the twin plane normal and the shear direction of the two active twin systems in Figure *1*b+c gave rise to suspicion of differences. The MD simulations clearly show that the differences lie in the stresses causing twin formation.

In Figure *5*a, the RSS values based on the Hamilton stress field are plotted as a function of the rotation angle $\varphi$ for the single crystal with ND $[0\bar{1}\bar{1}]$. The same procedure as for the single crystal with ND $[00\bar{1}]$ was applied. The maximum RSS value was determined for all SD possibilities on the single crystal with ND $[0\bar{1}\bar{1}]$. The different SDs are again represented by the rotation angle $\varphi$. A detailed diagram for all slip and twin systems is given in Fig. S3b+d. An angle of $\varphi = 0°$ corresponds to a SD of $[100]$, 90° to $[0\bar{1}1]$ and 145° to $[\bar{2}11]$. Experimentally, twins were only observed in a single crystal tested with ND $[0\bar{1}\bar{1}]$ SD $[100]$. Here, the RSS has the lowest values on the twin systems. The geometrically identified twin system $[\bar{1}11]$ $[2\bar{1}\bar{1}]$ does not even have the highest RSS of the twelve possible ones in SD $[100]$. Therefore and at first glance, the RSS based on the Hamilton stress field does not reflect the experimental results on the single crystal with ND $[0\bar{1}\bar{1}]$. However and based on the MD simulations, the location of twin formation is in front of the sphere. For this reason, the lateral position of the resolved shear stress maximum was investigated. In Figure *5*b, the twin systems with their maxima in front of the contact area for the different SDs are presented. The corresponding values of the RSS maxima demonstrate that the geometrically identified system in Figure *1*c in SD $[100]$ has the highest value. The stress distribution map of $(\bar{1}\bar{1}1)$ $[2\bar{1}\bar{1}]$ in SD $[100]$ in Figure *5*c also reveals a positive RSS at positive x-values, which corresponds to a position in front of the contact point. The RSS on the slip systems are higher compared to the twin systems and also higher in comparison to the single crystal with ND $[00\bar{1}]$. Based on this, pronounced dislocation-mediated plasticity is expected. Dislocation glide is expected to decrease the stresses acting within the material. Twin systems with their RSS maximum the furthest in front of the sphere will be least affected. In turn, the RSS on the other twin systems will be decreased more strongly by dislocation motion. The sole remaining point of possible criticism of this train of thought is that the RSS on the twin system $(\bar{1}\bar{1}1)$ $[2\bar{1}\bar{1}]$ does not exceed the critical RSS for twinning published in literature. This is true and will need further deliberations to understand or perhaps rather points to existing weaknesses inherent to our approach, which we might be able to rectify by using a more realistic model for the stress field under the moving sphere, which also need to consider plasticity.

This approach successfully explains that twinning is caused by $\sigma_{\text{SD}}$ on the single crystal with ND $[0\bar{1}\bar{1}]$ in front of the indenter. With this additional knowledge about the location of twin formation, the RSS values based on the Hamilton stress field are in full agreement with the results of the MD simulations.

This highlights the usefulness of this approach to evaluate and develop a fundamental understanding of the deformation mechanisms acting during a tribological load.

In this work, dry single-trace tribological experiments were conducted with SiC spheres sliding on single-crystalline CoCrFeMnNi. Combinations of crystallographic directions parallel to SD and ND were systematically investigated with respect to the activation of deformation twinning. These experiments were performed to demonstrate the potential use of deformation twinning as a probe for the stress field under a sliding sphere in a tribological contact. We gained important insight into which stress component causes twinning and its location with respect to the centre of the contact. Neither the crystal direction parallel to ND nor SD alone are decisive for twinning. The experimentally observed twins have in common that their twin plane normals and shear directions are all within the SD-ND plane. This geometric criterion for deformation twinning in a sliding contact is fulfilled for crystallographic orientations with a ⟨110⟩ direction parallel to TD, allowing the conclusion that crystal orientations with ⟨110⟩ parallel to TD exhibit twin formation. As twin systems are associated with maximum resolved shear stresses, the linear elastic Hamilton stress field approximation was employed for assessing the impact of the stress distribution. In the case of single crystals with ND $[00\bar{1}]$, the maximum resolved shear stresses for twinning and dislocation slip fully agree with the experimentally observed activation or absence of deformation twinning. To additionally capture atomistic details of deformation twinning under a tribological load, molecular dynamics simulations were carried out. The stresses necessary to activate twinning and the locations were this process takes place are identified to be different between the NDs $[00\bar{1}]$ and $[0\bar{1}\bar{1}]$. Twins for the combination of ND $[00\bar{1}]$ and SD $[\bar{1}10]$ were identified to form underneath the indenter and close to the leading edge of the sliding sphere. Here, twinning is mainly determined by $\sigma_{\mathrm{ND}}$. In contrast, the twins for a combination of ND $[0\bar{1}\bar{1}]$ and SD $[100]$ form in front of the sphere; a process which is mainly governed by compressive $\sigma_{\mathrm{SD}}$ stresses. Taking this finding about the twinning location into account, the Hamilton based analysis then also works in the case of ND $[0\bar{1}\bar{1}]$. This combination of experimental, analytical and simulative results fully demonstrate that deformation twinning, with its unique geometrical rules, is an ideal probe for the complex stress field associated with tribological contacts. Within reason, statements about the location and magnitude of specific stress field components can be made.

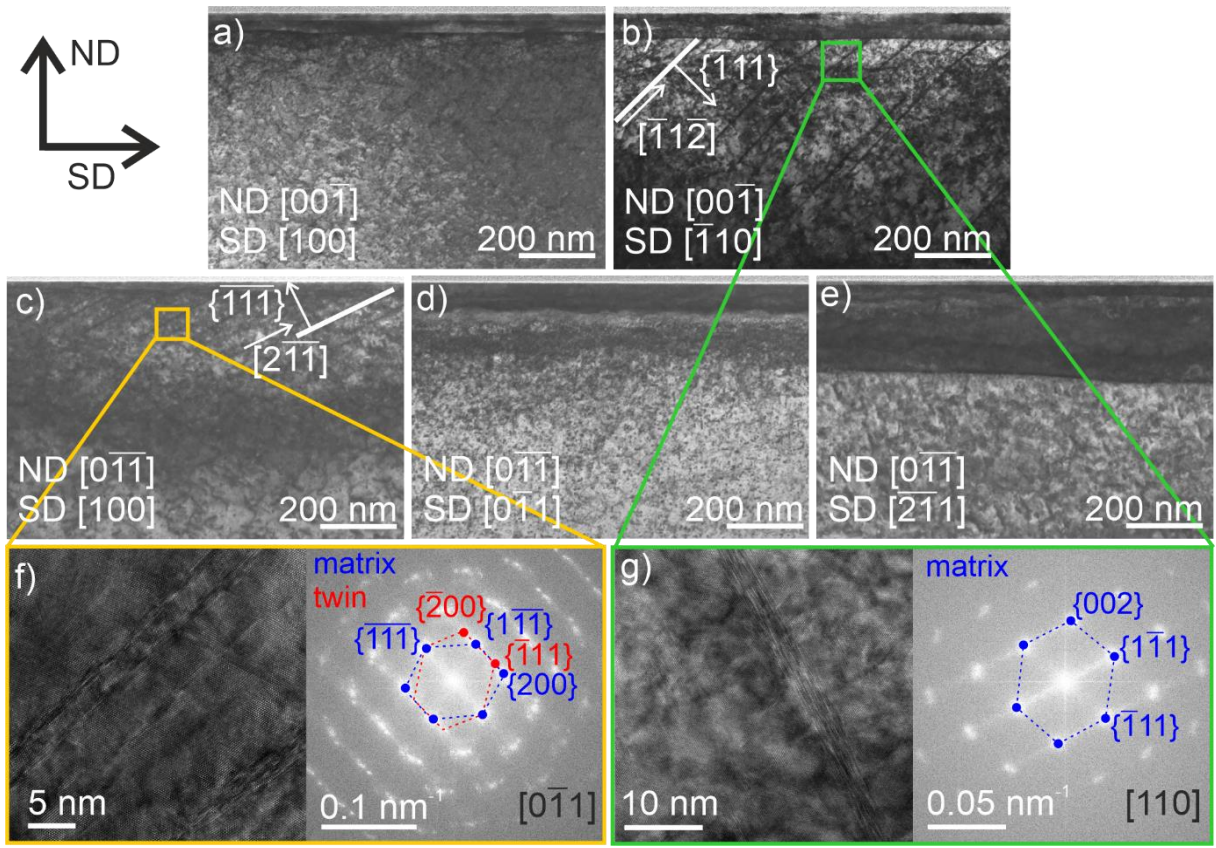

Figure 1. Scanning transmission electron microscopy (STEM) images of the deformation layers for the different crystal orientations chosen, as well as HR-TEM images of the twins. STEM images of the crystal orientation in a) ND [00$\bar{1}$] SD [100], b) ND [00$\bar{1}$] SD [$\bar{1}$10], c) ND [0$\bar{1}\bar{1}$] SD [100], d) ND [0$\bar{1}\bar{1}$] SD [0$\bar{1}$1] and e) ND [0$\bar{1}\bar{1}$] SD [$\bar{2}$11]; f) HR-TEM image and corresponding FFT of the lines inclined in SD in c), g) HR-TEM image and corresponding FFT of the lines inclined in SD in b). In b) and c), twin systems are presented by the twin plane normal and Shockley partial dislocation. In the STEM images, SD always is from left to right. Surfaces are covered by protective platinum layers.

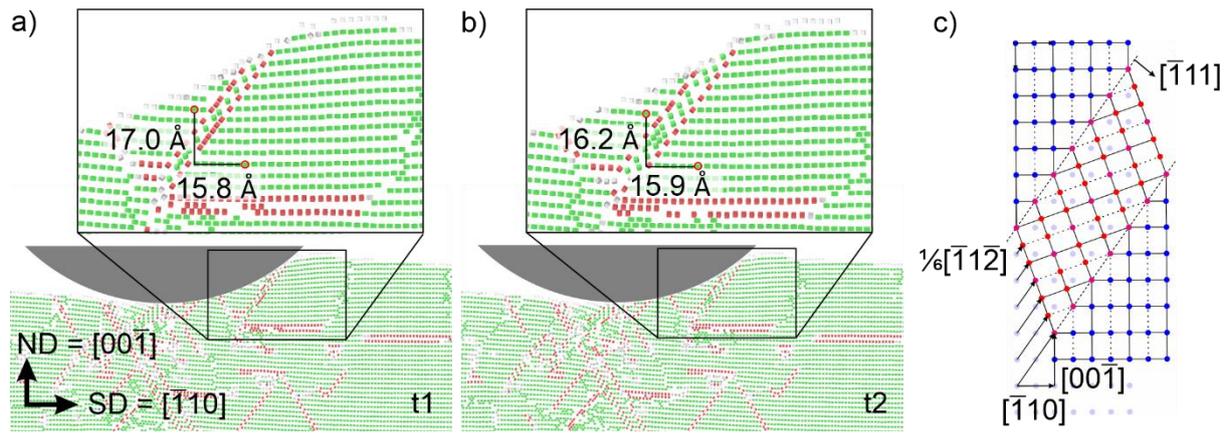

Figure 2. Molecular dynamics (MD) simulation of the microstructural evolution for ND $[00\bar{1}]$ SD $[\bar{1}10]$. Two representative timesteps are presented in a) and b), before and after twin thickening respectively. The insets zoom in to the twin system of interest. Atoms on the left and right side of the twin are marked in a) by red circles, and the distance in ND and SD was calculated based on the atomic positions in the MD simulation. The atom ID allowed to mark the same atoms in b) after twin thickening. c) presents a schematic drawing of the activated twin system. Here, blue circles are matrix atoms, red circles are twin atoms and light blue circles are the atom position before twinning. The SD is from left to right.

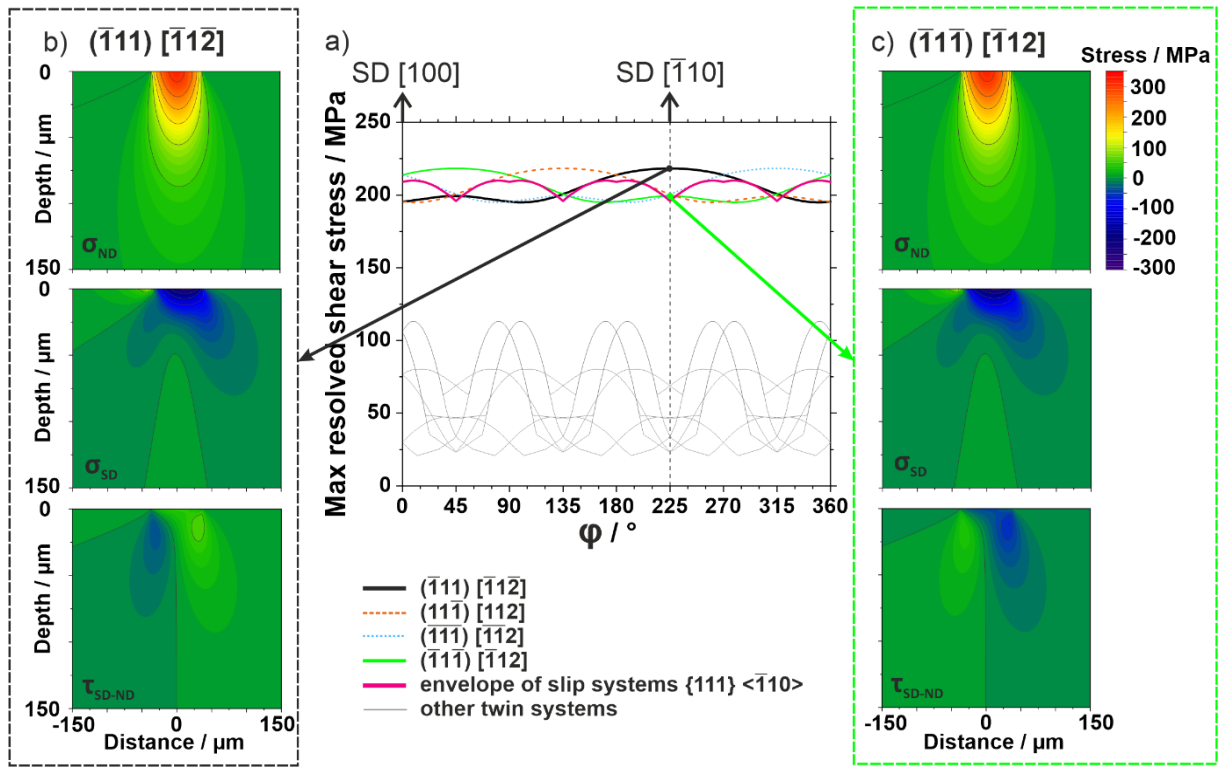

Figure 3. Resolved shear stress calculations employing the Hamilton stress field analysis for a single crystal with ND $[00\bar{1}]$. a) Maximum resolved shear stress as a function of the rotation angle of SD around the fixed ND, designated $\varphi$. An angle $\varphi$ equals to 0° corresponds to a SD of $[100]$, and $\varphi$ equal to 225° to $[\bar{1}10]$. The resolved shear stress distribution maps of various stress components of the twin systems $(\bar{1}11)\,[\bar{1}1\bar{2}]$ in b) and $(\bar{1}1\bar{1})\,[\bar{1}12]$ in c) at ND $[00\bar{1}]$ SD $[\bar{1}10]$ are presented. The angle – thus sliding direction – and twin systems are marked in a). Additional stress distribution maps are presented in Figure S4. The contact point of the two bodies making up the tribological system is at a distance of 0 µm. Positions in front of the contact point with respect to SD take positive x-values.

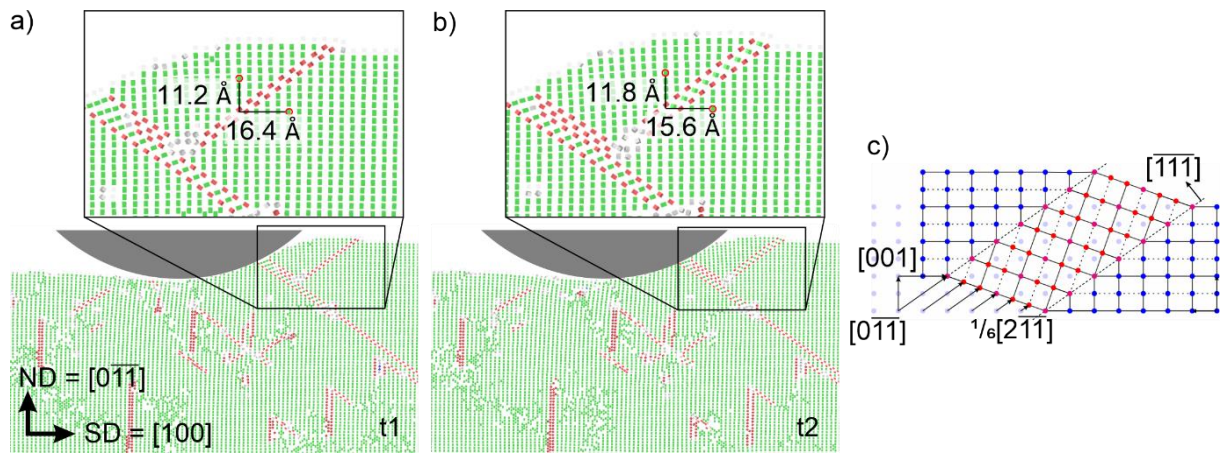

Figure 4. MD simulations of the microstructural evolution in ND $[0\bar{1}\bar{1}]$ and SD $[100]$. Two representative time steps are presented in a) and b), before and after twin thickening respectively. The insets zoom in to the twin system of interest. Atoms on the left and right side of the twin have been marked in a) by red circles, and the distance in ND and SD was calculated based on the atom positions in the MD simulations. The atom ID allowed to mark the same atoms in b) after twin thickening. c) presents a schematic drawing of the activated twin system. Here, blue circles are matrix atoms, red circles are twin atoms and light blue circles are the atom position before twinning. The SD is from left to right.

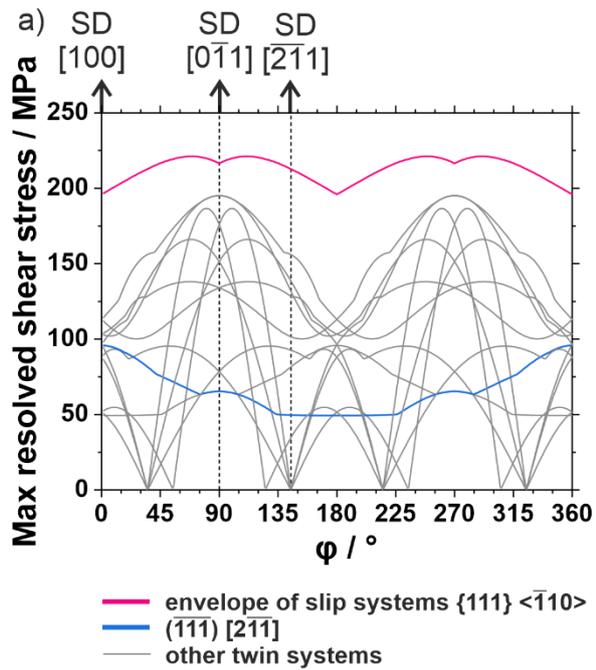

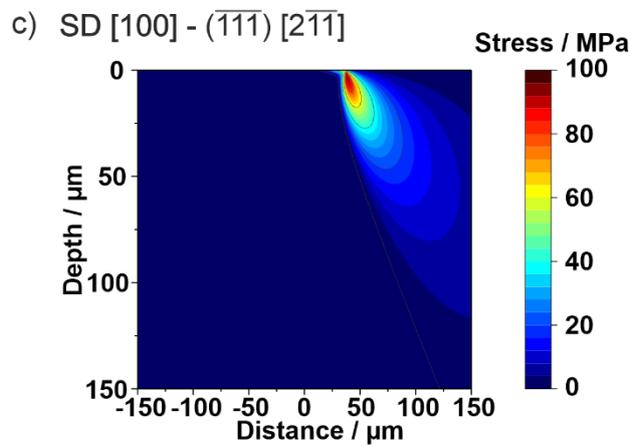

| b) | Twin system | max. RSS |
|---|---|---|
| SD [100] | $(\bar{1}\bar{1}1)\,[2\bar{1}\bar{1}]$ | 96 MPa |
| | $(\bar{1}11)\,[211]$ | 49 MPa |
| | $(\bar{1}1\bar{1})\,[21\bar{1}]$ | 52 MPa |
| | $(11\bar{1})\,[\bar{2}1\bar{1}]$ | 51 MPa |
| SD [0$\bar{1}$1] | $(\bar{1}\bar{1}1)\,[2\bar{1}\bar{1}]$ | 65 MPa |
| | $(\bar{1}11)\,[211]$ | 65 MPa |
| SD [$\bar{2}$11] | $(\bar{1}11)\,[21\bar{1}]$ | 1 MPa |
| | $(11\bar{1})\,[\bar{2}1\bar{1}]$ | 35 MPa |
| | $(\bar{1}11)\,[211]$ | 81 MPa |
| | $(\bar{1}\bar{1}1)\,[2\bar{1}\bar{1}]$ | 50 MPa |

Figure 5. Resolved shear stress calculations employing the Hamilton stress field analysis for a single crystal with ND [$0\bar{1}\bar{1}$]. a) Maximum resolved shear stress as a function of the rotation angle of SD around the fixed ND, designated $\varphi$. An angle of $\varphi$ = 0° corresponds to a SD of [100], $\varphi$ = 90° to [$0\bar{1}1$] and $\varphi$ = 145° to [$\bar{2}11$]; b) table of the twin systems with a maximum in resolved shear stress the furthest in front of the sphere and the corresponding maximum resolved shear stress for the experimentally investigated SDs. c) Resolved shear stress distribution map for the full stress field of the twin system $(\bar{1}\bar{1}1)$ [$2\bar{1}\bar{1}$] in SD [100].

# Methods

## Experimental details

CoCrFeMnNi single crystals were grown by means of the Bridgman technique. An alumina crucible of 20 mm diameter with a tip shaped bottom was charged with an equiatomic melt. The melt was pre-manufactured by levitation melting in a Cu cold crucible out of Co, Cr, Fe, Mn and Ni with purities of 3N+, 5N, 4N, 4N and 4N5, respectively. The crucible was placed on a water-cooled movable rod in the Bridgman furnace, which was held at 200 mbar argon atmosphere and a constant temperature of 1340 °C. The growth was carried out by withdrawing the crucible vertically out of the hot zone at a velocity of 50 mm/h.

The final crystal was investigated by Laue diffractometry, which demonstrated that it consisted of a single grain (except for two mm-sized surface grains, which could easily be removed) of about 6.5 cm$^3$. Optical microscopy of cuts perpendicular to the growth direction showed the presence of a consistent dendrite pattern with a primary branch diameter of 0.2 to 0.4 mm over the entire surface.

In order to homogenize the material and remove the dendrites, the single crystals were heat treated in a tube furnace at 1200 °C for 72 h. The samples were mechanically ground up to grit #4000 and polished with 3 µm and 1 µm diamond suspensions (Cloeren Technology GmbH, Wegberg, Germany) for at least eight minutes each. To achieve a defect-free surface, the samples were electropolished prior to the tribological experiments with an electrolyte from perchloric acid and methanol in a ratio of 1:9. The electrolyte was used at RT for the single crystal with ND $[00\bar{1}]$ and cooled in a fridge for ND $[0\bar{1}\bar{1}]$. After each preparation step, the samples were cleaned in an ultrasonic bath of isopropanol.

The tribological experiments were conducted with a self-constructed tribometer[22]. One important characteristic of this tribometer is a rotational stage which can be used to precisely orient the sample to the desired SDs. To adjust the atmosphere, the tribometer is encapsulated in a box of PMMA. All further experimental details are given in the main text.

TEM foils were cut with a dual beam scanning electron microscope (SEM), focused ion beam (FIB) (Helios NanoLabTM DualBeamTM 650, ThermoFisher, Hillsboro, USA). The scanning transmission electron microscopy (STEM) images were taken using the same microscope with 30 kV and 100 pA and a STEM detector in bright field conditions. The high resolution-TEM images were acquired with a double corrected Themis Z (ThermoFisher, Hillsboro, USA) operated at 300 kV equipped with a OneView CCD camera (Gatan Inc, Pleasanton, USA). For all investigated TEM foils, the zone axis was $\langle 110 \rangle$.

## MD simulation

All MD simulations were carried out using the open-source MD code LAMMPS[61], orientation analyses and visualization were performed using OVITO[54], while pre- and post-processing were done in MATLAB. An fcc single crystal with a lattice constant of 3.609 Å was rotated by 45° around the system's z-axis such that ND $[00\bar{1}]$ and SD $[\bar{1}10]$, and once around the x-axis such that ND was $[0\bar{1}\bar{1}]$ and SD $[100]$. The total crystal dimensions were 80x40x40 nm³, resulting in two systems of 10.9m atoms for each simulation. CoCrFeMnNi crystals were generated via a random substitution of the atom types such that every element constitutes exactly 20% of the final system. The atomic interactions were governed by the Gröger potential[62], which is effectively a collection of spline potentials for all pairwise combinations of elements in the alloy. The systems were fitted with two layers of 0.35 nm thickness each at the bottom (near z=0 nm), the lower one being kept rigid, and the other one acting as a heat sink set to 300 K that emulates the thermodynamic response of the bulk material via a fast Langevin thermostat ($\lambda$=0.5 ps). The remainder of each system was thermostated with a slower Langevin thermostat ($\lambda$=7 ps) to approximate the low thermal conductivity of CoCrFeMnNi, effectively implementing an electron-phonon coupling approach[63]. Both thermostats acted only in the direction perpendicular to the loading and scratching directions so as not to interfere with these external constraints. Periodic boundary conditions were applied along the lateral (x-y) dimensions of the systems.

During the simulation runs, a purely repulsive spherical indenter (force constant: 10 eV/Å³, radius: 20 nm) was moved into the sample surface at $v_x$ = 2.5 m/s and $v_z$ = –1 m/s until the scratch depth reached a value of 2.5 nm. After that, the vertical indenter position was kept constant, and only its displacement in x-direction progressed up to a total scratch length of 50 nm. The standard output interval for snapshots of the system geometries was 80 ps, and high-resolution output at an interval of 1 ps was additionally produced where required.

# References

1. Holmberg, K. & Erdemir, A. The impact of tribology on energy use and CO2 emission globally and in combustion engine and electric cars. *Tribology International* **135**, 389–396 (2019).

2. Blau, P. J. Interpretations of the friction and wear break-in behavior of metals in sliding contact. *Wear* **71**, 29–43 (1981).

3. Rigney, D. A. & Glaeser, W. A. The significance of near surface microstructure in the wear process. *Wear* **46**, 241–250 (1978).

4. Rigney, D. A. & Hirth, J. P. Plastic deformation and sliding friction of metals. *Wear* **53**, 345–370 (1979).

5. Heilmann, P., Clark, W. A. T. & Rigney, D. A. Orientation determination of subsurface cells generated by sliding. *Acta Metallurgica* **31**, 1293–1305 (1983).

6. Zum Gahr, K.-H. *Microstructure and Wear of Meaterials*. (Elsevier Science Publishers B. V., 1987).

7. Argibay, N., Chandross, M., Cheng, S. & Michael, J. R. Linking microstructural evolution and macro-scale friction behavior in metals. *J Mater Sci* **52**, 2780–2799 (2017).

8. Xia, W., Dehm, G. & Brinckmann, S. Investigation of single asperity wear at the microscale in an austenitic steel. *Wear* **452–453**, 203289 (2020).

9. Patil, P., Lee, S., Dehm, G. & Brinckmann, S. Influence of crystal orientation on twinning in austenitic stainless-steel during single micro-asperity tribology and nanoindentation. *Wear* **504–505**, 204403 (2022).

10. Tsuya, Y. The anisotropy of the coefficient of friction and plastic deformation in copper single crystals. *Wear* **14**, 309–322 (1969).

11. Laube, S. *et al.* Solid solution strengthening and deformation behavior of single-phase Cu-base alloys under tribological load. *Acta Materialia* **185**, 300–308 (2020).

12. Liu, Z., Messer-Hannemann, P., Laube, S. & Greiner, C. Tribological performance and microstructural evolution of α-brass alloys as a function of zinc concentration. *Friction* **8**, 1117–1136 (2020).

13. Eder, S. J., Cihak-Bayr, U., Gachot, C. & Rodríguez Ripoll, M. Interfacial microstructure evolution due to strain path changes in sliding contacts. *ACS Appl. Mater. Interfaces* **10**, 24288–24301 (2018).

14. Yao, B., Han, Z. & Lu, K. Dry sliding tribological properties and subsurface structure of nanostructured copper at liquid nitrogen temperature. *Wear* **301**, 608–614 (2013).

15. Rau, J. S., Schmidt, O., Schneider, R., Debastiani, R. & Greiner, C. Three Regimes in the Tribo-Oxidation of High Purity Copper at Temperatures of up to 150 °C. *Advanced Engineering Materials* **24**, 2200518 (2022).

16. Rynio, C., Hattendorf, H., Klöwer, J. & Eggeler, G. The evolution of tribolayers during high temperature sliding wear. *Wear* **315**, 1–10 (2014).

17. Chen, X., Schneider, R., Gumbsch, P. & Greiner, C. Microstructure evolution and deformation mechanisms during high rate and cryogenic sliding of copper. *Acta Materialia* **161**, 138–149 (2018).

18. Eder, S. J., Grützmacher, P. G., Rodríguez Ripoll, M., Gachot, C. & Dini, D. Does speed kill or make friction better?—Designing materials for high velocity sliding. *Applied Materials Today* **29**, 101588 (2022).

19. Emge, A., Karthikeyan, S., Kim, H. J. & Rigney, D. A. The effect of sliding velocity on the tribological behavior of copper. *Wear* **263**, 614–618 (2007).

20. Dollmann, A. *et al.* Dislocation-mediated and twinning-induced plasticity of CoCrFeMnNi in varying tribological loading scenarios. *J Mater Sci* **57**, 17448–17461 (2022).

21. Prasad, S. V., Michael, J. R. & Christenson, T. R. EBSD studies on wear-induced subsurface regions in LIGA nickel. *Scripta Materialia* **48**, 255–260 (2003).

22. Haug, C., Molodov, D., Gumbsch, P. & Greiner, C. Tribologically induced crystal rotation kinematics revealed by electron backscatter diffraction. *Acta Materialia* **225**, 117566 (2022).

23. Cai, W., Bellon, P. & Beaudoin, A. J. Probing the subsurface lattice rotation dynamics in bronze after sliding wear. *Scripta Materialia* **172**, 6–11 (2019).

24. Cai, W. & Bellon, P. Microstructural self-organization triggered by twin boundaries during dry sliding wear. *Acta Materialia* **60**, 6673–6684 (2012).

25. Greiner, C., Liu, Z., Strassberger, L. & Gumbsch, P. Sequence of stages in the microstructure evolution in copper under mild reciprocating tribological loading. *ACS applied materials & interfaces* **8**, 15809–15819 (2016).

26. Prasad, S. V., Michael, J. R., Battaile, C. C., Majumdar, B. S. & Kotula, P. G. Tribology of single crystal nickel: Interplay of crystallography, microstructural evolution, and friction. *Wear* **458**, 203320 (2020).

27. Greiner, C., Liu, Z., Schneider, R., Pastewka, L. & Gumbsch, P. The origin of surface microstructure evolution in sliding friction. *Scripta Materialia* **153**, 63–67 (2018).

28. Xia, W., Patil, P. P., Liu, C., Dehm, G. & Brinckmann, S. A novel microwall sliding test uncovering the origin of grain refined tribolayers. *Acta Materialia* **246**, 118670 (2023).

29. Haug, C. *et al.* Early deformation mechanisms in the shear affected region underneath a copper sliding contact. *Nat Commun* **11**, 1–8 (2020).

30. Ruebeling, F. *et al.* Normal load and counter body size influence the initiation of microstructural discontinuities in Copper during sliding. *ACS Appl. Mater. Interfaces* acsami.0c19736 (2021) doi:10.1021/acsami.0c19736.

31. Hughes, D. A. & Hansen, N. Graded nanostructures produced by sliding and exhibiting universal behavior. *Physical Review Letters* **87**, 135503 (2001).

32. Brinckmann, S. & Dehm, G. Nanotribology in austenite: Plastic plowing and crack formation. *Wear* **338**, 436–440 (2015).

33. Gagel, J., Weygand, D. & Gumbsch, P. Discrete Dislocation Dynamics simulations of dislocation transport during sliding. *Acta Materialia* **156**, 215–227 (2018).

34. Zhu, J. *et al.* Coupled effect of scratching direction and speed on nano-scratching behavior of single crystalline copper. *Tribology International* **150**, 106385 (2020).

35. Zhu, J., Li, X., Zhou, Q. & Aghababaei, R. On the anisotropic scratching behavior of single crystalline copper at nanoscale. *Tribology International* **175**, 107794 (2022).

36. Kareer, A., Tarleton, E., Hardie, C., Hainsworth, S. V. & Wilkinson, A. Scratching the surface: Elastic rotations beneath nanoscratch and nanoindentation tests. *arXiv:2006.12554 [cond-mat]* (2020).

37. Hamilton, G. M. Explicit Equations for the Stresses beneath a Sliding Spherical Contact. *Proceedings of the Institution of Mechanical Engineers, Part C: Journal of Mechanical Engineering Science* **197**, 53–59 (1983).

38. Mahajan, S. & Williams, D. F. Deformation Twinning in Metals and Alloys. *International Metallurgical Reviews* **18**, 43–61 (1973).

39. Rainforth, W. M., Stevens, R. & Nutting, J. Deformation structures induced by sliding contact. *Philosophical Magazine A* **66**, 621–641 (1992).

40. Lychagin, D. V. *et al.* Dry sliding of Hadfield steel single crystal oriented to deformation by slip and twinning: Deformation, wear, and acoustic emission characterization. *Tribology International* **119**, 1–18 (2018).

41. Dollmann, A., Kauffmann, A., Heilmaier, M., Haug, C. & Greiner, C. Microstructural changes in CoCrFeMnNi under mild tribological load. *J Mater Sci* **55**, 12353 (2020).

42. Okamoto, N. L. *et al.* Size effect, critical resolved shear stress, stacking fault energy, and solid solution strengthening in the CrMnFeCoNi high-entropy alloy. *Scientific Reports* **6**, 35863 (2016).

43. Karaman, I., Sehitoglu, H., Gall, K., Chumlyakov, Y. I. & Maier, H. J. Deformation of single crystal Hadfield steel by twinning and slip. *Acta Materialia* **48**, 1345–1359 (2000).

44. Edington, J. W. Electron Diffraction in the Electron Microscope. in *Electron Diffraction in the Electron Microscope* (ed. Edington, J. W.) 1–77 (Macmillan Education UK, 1975). doi:10.1007/978-1-349-02595-4_1.

45. Otto, F. *et al.* The influences of temperature and microstructure on the tensile properties of a CoCrFeMnNi high-entropy alloy. *Acta Materialia* **61**, 5743–5755 (2013).

46. Martin, S. *et al.* Deformation behaviour of TWIP steels: Constitutive modelling informed by local and integral experimental methods used in concert. *Materials Characterization* **184**, 111667 (2022).

47. Venables, J. A. Deformation Twinning in Face-Centred Cubic Metals. *Philosophical Magazine* **6**, 379–396 (1961).

48. Cohen, J. B. & Weertman, J. A dislocation model for twinning in f.c.c. metals. *Acta Metallurgica* **11**, 996–998 (1963).

49. Fujita, H. & Mori, T. A formation mechanism of mechanical twins in F.C.C. Metals. *Scripta Metallurgica* **9**, 631–636 (1975).

50. Kang, S., Jung, Y.-S., Yoo, B.-G., Jang, J. & Lee, Y.-K. Orientation-dependent indentation modulus and yielding in a high Mn twinning-induced plasticity steel. *Materials Science and Engineering: A* **532**, 500–504 (2012).

51. Seo, E. J. *et al.* Micro-plasticity of medium Mn austenitic steel: Perfect dislocation plasticity and deformation twinning. *Acta Materialia* **135**, 112–123 (2017).

52. Christian, J. W. & Mahajan, S. Deformation twinning. *Progress in Materials Science* **39**, 1–157 (1995).

53. Dollmann, A. *et al.* Temporal sequence of deformation twinning in CoCrNi under tribological load. *Scripta Materialia* **229**, 115378 (2023).

54. Stukowski, A. Visualization and analysis of atomistic simulation data with OVITO–the Open Visualization Tool. *Modelling Simul. Mater. Sci. Eng.* **18**, 015012 (2009).

55. Rafaja, D., Ullrich, C., Motylenko, M. & Martin, S. Microstructure Aspects of the Deformation Mechanisms in Metastable Austenitic Steels. in *Austenitic TRIP/TWIP Steels and Steel-Zirconia Composites: Design of Tough, Transformation-Strengthened Composites and Structures* (eds. Biermann, H. & Aneziris, C. G.) 325–377 (Springer International Publishing, 2020). doi:10.1007/978-3-030-42603-3_11.

56. Kawamura, M. *et al.* Plastic deformation of single crystals of the equiatomic Cr–Mn–Fe–Co–Ni high-entropy alloy in tension and compression from 10 K to 1273 K. *Acta Materialia* **203**, 116454 (2021).

57. Abuzaid, W. & Sehitoglu, H. Critical resolved shear stress for slip and twin nucleation in single crystalline FeNiCoCrMn high entropy alloy. *Materials Characterization* **129**, 288–299 (2017).

58. Kireeva, I. V., Chumlyakov, Yu. I., Vyrodova, A. V., Pobedennaya, Z. V. & Karaman, I. Effect of twinning on the orientation dependence of mechanical behaviour and fracture in single crystals of the equiatomic CoCrFeMnNi high-entropy alloy at 77K. *Materials Science and Engineering: A* **784**, 139315 (2020).

59. Patriarca, L., Ojha, A., Sehitoglu, H. & Chumlyakov, Y. I. Slip nucleation in single crystal FeNiCoCrMn high entropy alloy. *Scripta Materialia* **112**, 54–57 (2016).

60. Wagner, C. & Laplanche, G. Effect of grain size on critical twinning stress and work hardening behavior in the equiatomic CrMnFeCoNi high-entropy alloy. *International Journal of Plasticity* **166**, 103651 (2023).

61. Thompson, A. P. *et al.* LAMMPS - a flexible simulation tool for particle-based materials modeling at the atomic, meso, and continuum scales. *Computer Physics Communications* **271**, 108171 (2022).

62. Gröger, R., Vitek, V. & Dlouhý, A. Effective pair potential for random fcc CoCrFeMnNi alloys. *Modelling Simul. Mater. Sci. Eng.* **28**, 075006 (2020).

63. Eder, S. J., Cihak-Bayr, U., Bianchi, D., Feldbauer, G. & Betz, G. Thermostat Influence on the Structural Development and Material Removal during Abrasion of Nanocrystalline Ferrite. *ACS applied materials & interfaces* **9**, 13713–13725 (2017).

## Acknowledgements

CG and AD thank the German Research Foundation (DFG) for funding under Project GR 4174/5-1. SJE acknowledges the Austrian COMET-Program (Project K2 InTribology1, no. 872176). Computational results were obtained using the Vienna Scientific Cluster (VSC). VT acknowledges PhD funding by the German Academic Exchange Service (DAAD). MF thanks the Priority Program 'Compositionally Complex Alloys – High Entropy Alloys (CCA-HEA)' of DFG for funding under the project number FE 571/4-2. Thanks also to Martin Heilmaier for his support and interest in this work.

## Data availability

The raw data of this study are available under the DOI 10.35097/1732 and from the corresponding author upon request.

Supplementary Information

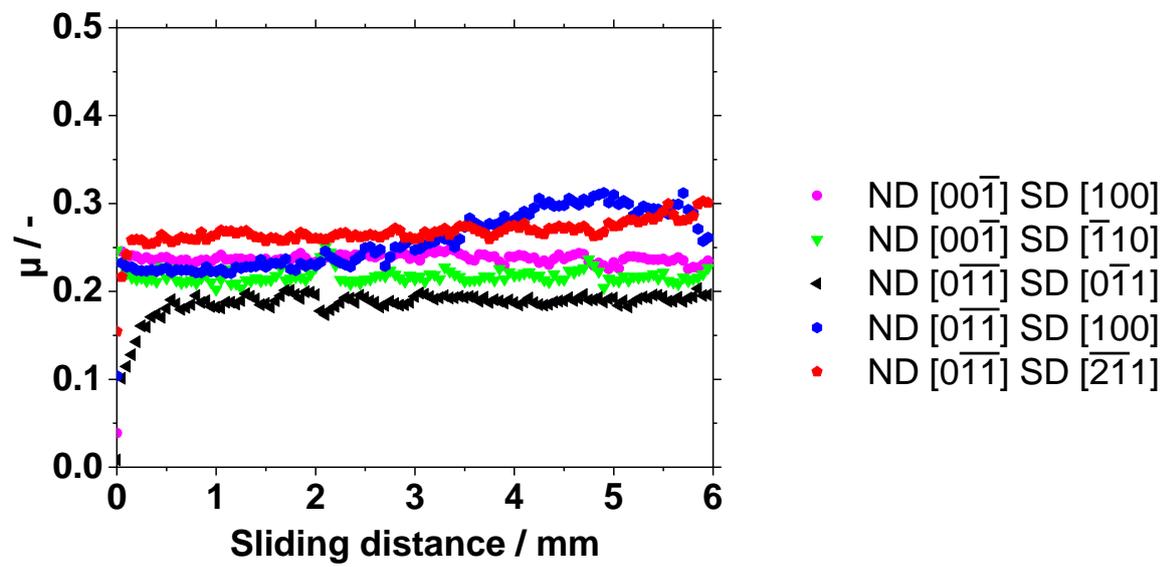

Figure S1. Friction coefficient µ as a function of the sliding distance for all crystal orientations and sliding directions investigated.

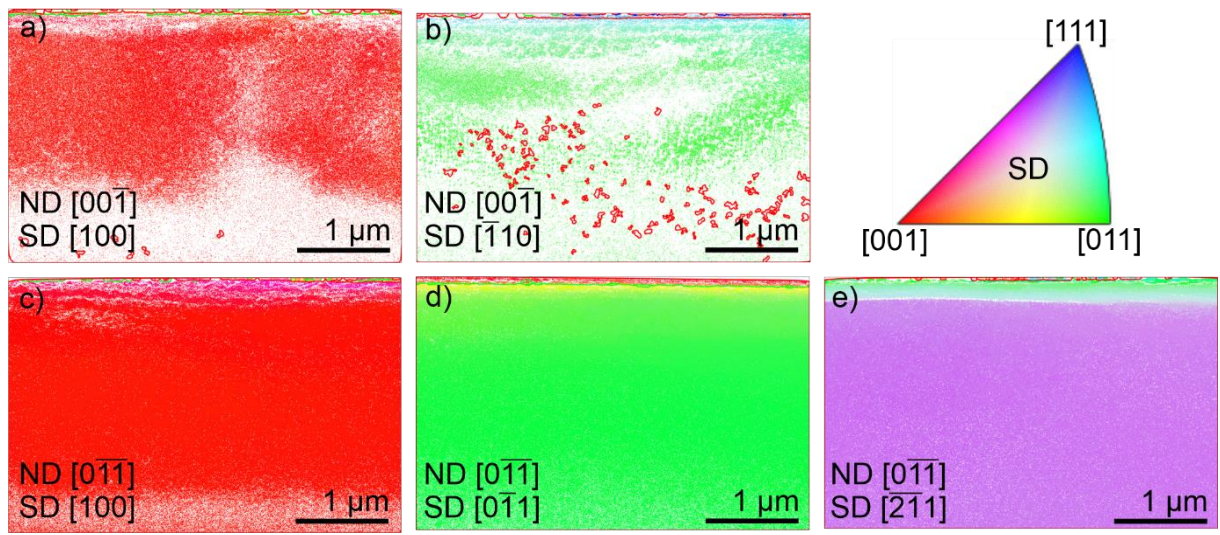

Figure S2. Transmission Kikuchi diffraction (TKD) measurements. a) for ND $[00\bar{1}]$ SD $[100]$, b) ND $[00\bar{1}]$ SD $[\bar{1}10]$, c) ND $[0\bar{1}\bar{1}]$ SD $[100]$, d) ND $[0\bar{1}\bar{1}]$ SD $[0\bar{1}1]$ and e) ND $[0\bar{1}\bar{1}]$ SD $[\bar{2}\bar{1}1]$. The color-coding in SD is given in the inverse pole figure. Green lines represent small angle grain boundaries (3-15°), red lines high angle grain boundaries (> 15°) and blue lines twins. The sliding direction is from left to right.

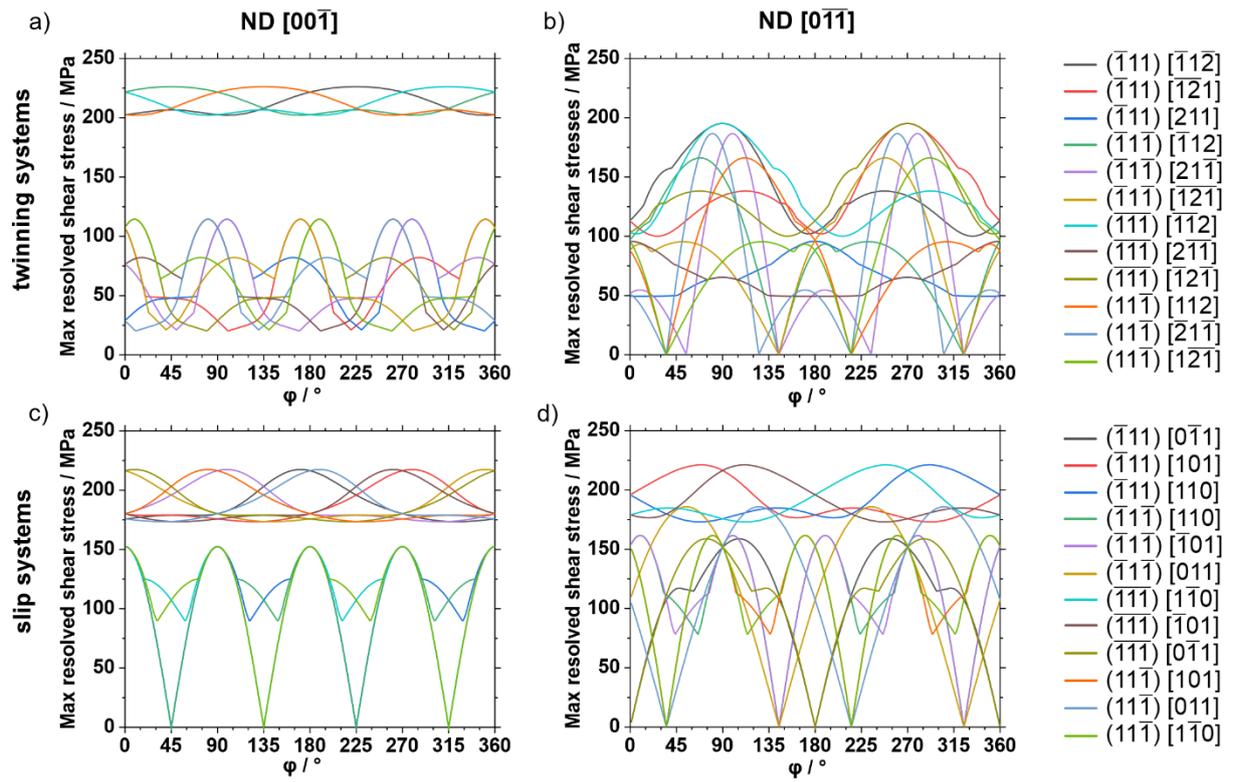

Figure S3. Maximum resolved shear stresses as a function of the rotation angle around ND. Maximum RSS on the twin systems with a) ND $[00\bar{1}]$ and b) ND $[0\bar{1}\bar{1}]$. Absolute maximum RSS on the slip systems with c) ND $[00\bar{1}]$ and d) ND $[0\bar{1}\bar{1}]$.

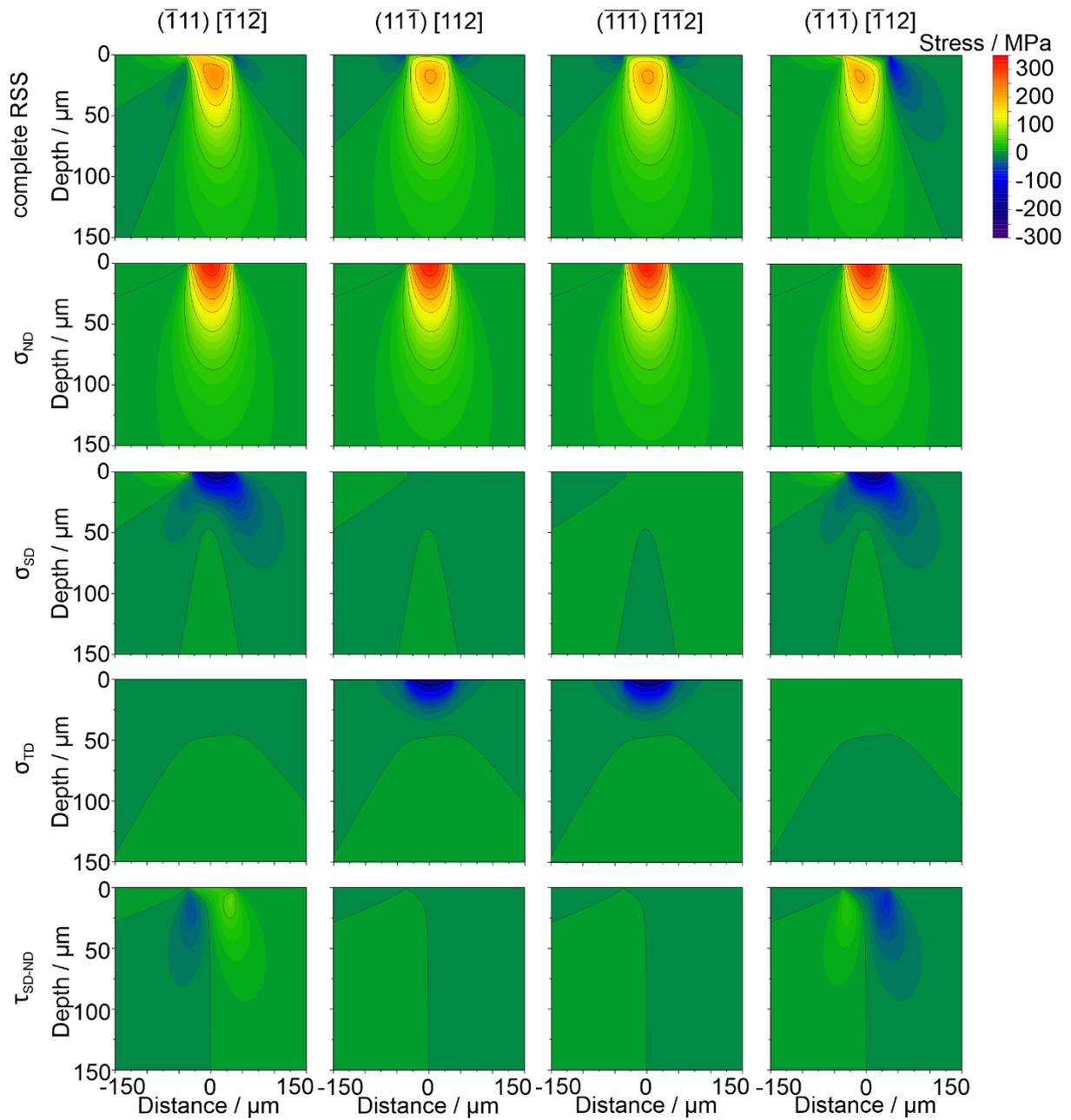

Figure S4. Stress distribution maps for the crystal orientation with ND $[00\bar{1}]$ SD $[\bar{1}10]$ for the four twin systems with the highest resolved shear stresses. The first column shows the results for $(\bar{1}11)\,[\bar{1}1\bar{2}]$, the second for $(11\bar{1})\,[112]$, the third for $(\overline{111})\,[\overline{11}2]$ and the fourth for $(\overline{11}1)\,[\bar{1}12]$. The first row shows the RSS calculated with the entire stress field, the second shows the RSS calculated with $\sigma_{\mathrm{ND}}$, the third the RSS calculated with $\sigma_{\mathrm{SD}}$, the fourth shows the RSS calculated with $\sigma_{\mathrm{TD}}$ and the fifth the RSS calculated with $\tau_{\mathrm{SD-ND}}$. The contact point between sample and sphere is at a distance of 0 μm. Positions with a positive x-value are in front of the indenter with respect to SD.